\documentclass[preprint,a4paper,12pt,3p]{article}
\usepackage[top=1in,bottom=1in,left=1in,right=1in]{geometry}

\usepackage{graphicx}

\usepackage{amssymb,amsmath,natbib,subfigure,hyperref}
\usepackage[misc,geometry]{ifsym}

\begin{document}

\title{Non-fixation in infinite potential}

\author{Song Xu\footnote{First Author, song.xu.sjtu@gmail.com}~, Shuyun Jiao, Pengyao Jiang, Bo Yuan\footnote{Corresponding Author, boyuan@sjtu.edu.cn}, Ping Ao\footnote{Corresponding Author, aoping@sjtu.edu.cn} \\ \normalsize{Shanghai Jiao Tong University, 200240, Shanghai, China}}

\maketitle

\begin{abstract}
Under the effects of strong genetic drift, it is highly probable to observe gene fixation or loss in a population, shown by divergent probability density functions, or infinite adaptive peaks on a landscape. It is then interesting to ask what such infinite peaks imply, with or without combining other biological factors (e.g. mutation and selection). We study the stochastic escape time from the generated infinite adaptive peaks, and show that Kramers' classical escape formula can be extended to the non-Gaussian distribution cases. The constructed landscape provides a global description for system's middle and long term behaviors, breaking the constraints in previous methods.
\end{abstract}

\section{Introduction}

Evolutionary processes, driven by deterministic and stochastic forces, can generate very complex phenomena in biology. One of the most typical and important issues is to model rare events and to calculate the transition probabilities between meta-stable states. Related problems have been referred in different contexts in literature. \cite{Wright1932} stressed that adaptation may be limited not by the rate toward local adaptive peaks but by the peak-to-peak transition rate, in his shifting-balance theory (SBT). \cite{Kimura1962} studied how the success or failure of a mutant gene depends on chance for all levels of selective dominance. \cite{Barton1987} stated that divergence of populations into different equilibria would reduce the fitness of hybrids. \cite{Gavrilets2003} reviewed how genetic barriers for gene flow are established and related it to biological speciation. Results on multiple adaptive peaks are also important for studying evolutionary robustness \citep{Ao2009}. The ideas and methodologies are also widely discussed outside biology \citep{Qian2005, Tyll2010}

Typically, the existence of a genetic barrier (or adaptive valley) suggests the separation of different evolutionary timescales. In chemistry, the famous Arrehnius formula estimates the separation factor by an exponential term of the barrier height (or valley depth) \citep{Gardiner1985}. It was latter systematically studied by \cite{Kramers1940} in thermally activated systems. In population genetics, however, random drift may cause problems for the application of the classical methods, such as generating infinitely high genetic barrier on a landscape, if can be properly defined. The questions here are: What is the biological meaning of these infinite peaks? Do they imply the fixation of an allele type? If not, what is the life time of such states? The classical escape formula would give biologically unexpected estimations here, and the correct results are often numerically computed or can be analytically analyzed under very special parameter settings \citep{Kimura1969, Lande1985, Joe2011}. In the present work, we will show how Kramers' classical formula can be extended to treat the infinite peaks, and that analytical estimations can be obtained.

\cite{Barton1987} proposed an eigenvalue method to calculate the escape rate in the diffusion model, which can in part deal with the divergent (infinite) peaks. When selection is very weak, however, their analytical estimation breaks down. We note that the essential problem comes from that their ``deterministic equilibrium" fails to describe the bi-stability of the system for certain cases. A proper framework should be used for analyzing the evolutionary dynamics. Another approach is more from the side of population genetics, referred as the calculation of ``rate of genetic substitution" \citep{Kimura1962, Gillespie1998}. It bypasses the problem of infinite peaks, but the results are not generally applicable for incalculable fixation probabilities. The present approach does not have this constraint and allows more complex dynamics, which provides a complete answer for the present bi-stability problems.

The present article is organized as follows: In Section 2 we introduce the 1-d diffusion process and defines a potential landscape. We then discuss the two typical timescales in the bi-stable models. In Section 3 we calculate the escape time from an infinite adaptive peak. We first study the mutation-drift model and then come to two types of selection. In Section 4 we compare our results to the previous efforts. We then discuss the obtained biological insights.

\section{Diffusion model and potential landscape}
\subsection{Wright-Fisher diffusion model}
The 1-d Wright-Fisher model considers the evolution of a diploid population at one locus. Assume that the generations are non-overlapping and the population size $N$ is big enough for the continuous diffusion approximation. Denote the interested pair of alleles as $A_1$ and $A_2$, with respective frequencies $x$ and $1-x$. Let $\rho(x,t)$ be the probability distribution that $A_1$ frequency is $x$ at time $t$. The diffusion equation for the continuous Wright-Fisher model is given by \citep{Kimura1964, Ewens2004, Blythe2007}:
\begin{equation}
\partial_t\rho(x,t)=\frac{1}{2}\partial^2_x\Big[V(x)\rho(x,t)\Big]- \partial_x\Big[M(x)\rho(x,t)\Big]~.
\label{Eq:2.1-1}
\end{equation}
$M(x)$ is the average change of $A_1$ frequency per generation, corresponding to the deterministic factors of the system. $V(x)$ is the variance of the stochastic factors. For example, under mutation and selection:
\begin{equation}
M(x) = -\mu x + \nu (1-x) + \dfrac{x(1-x)}{2\overline{\omega}}\dfrac{d\overline{\omega}}{dx}~, \label{Eq:2.1-7}
\end{equation}
where $\mu$ is the mutation rate from $A_{1}$ alleles to $A_{2}$ alleles, $\nu$ is that from $A_2$ to $A_1$; $\overline{\omega}$ gives the average fitness of the population. Under random genetic drift:
\begin{equation}
V(x) = x(1-x)/2N ~.
\label{Eq:2.1-8}
\end{equation}
$2N$ is the number of alleles in the interested locus in the diploid population.

\subsection{Potential landscape}
The equilibrium distribution of Eq.~(\ref{Eq:2.1-1}) can be easily obtained as:
\begin{equation}
\rho(x,\infty)= \dfrac{1}{V(x)}\exp\bigg[\int^x \dfrac{2M(y)}{V(y)}dy\bigg]\bigg/{Z}
= \exp\bigg[\int^x \dfrac{2M(y)-V'(y)}{V(y)}dy\bigg]\bigg/{Z} ~,
\label{Eq:2.1-5}
\end{equation}
where the normalization constant is given by
\begin{equation}
Z = \int^1_0 \exp\bigg[\int^x \dfrac{2M(y)-V'(y)}{V(y)}dy\bigg] dx~.
\label{Eq:2.1-6}
\end{equation}
Note that the form of Eq.~(\ref{Eq:2.1-5}) immediately suggests a potential function, or landscape, from the Boltzmann-Gibbs distribution \citep{Ao2005}
\begin{equation}
\Phi(x)= \int^x \dfrac{2M(y)-V'(y)}{V(y)}dy \doteq \int^x \frac{f(y)}{D(y)}dy~.
\label{Eq:2.1-4}
\end{equation}
Here we have defined a directed force $f(x)$ and an undirected diffusion term $D(x)$, which are closely related to the system's long-term dynamics:
\begin{align}
& f(x)=M(x)-V'(x)/2~, \label{Eq:2.1-3a} \\
& \epsilon D(x)=V(x)/2~. \label{Eq:2.1-3b}
\end{align}
If we rewrite Eq.~(\ref{Eq:2.1-1}) by substituting $M$ and $V$ with $f$ and $D$, we may obtain a desired symmetric form of Eq.~(\ref{Eq:2.1-1}), where the potential landscape can be directly read if the detailed-balance condition is satisfied \citep{Ao2005}. We can specify Eq.~(\ref{Eq:2.1-4}) in the Wright-Fisher model by considering Eqs.~(\ref{Eq:2.1-7}) and (\ref{Eq:2.1-8}):
\begin{equation}
\Phi(x)= -\ln x(1-x) + 4N\Big[\nu\ln x+\mu\ln(1-x)\Big]+ 2N\ln\overline{\omega} ~.
\label{Eq:2.1-9}
\end{equation}
This potential form has been mentioned several times in literature, such as by \cite{Burger2000}. With $\Phi(x)$, we may classify the Wright-Fisher processes under different parameters according to their long-term behaviors (Figure 1).

\subsection{Uphill and downhill movements}
We are interested in the bi-stable dynamics in the Wright-Fisher diffusion model. If starting from the simplest mutation-drift case, we have
\begin{equation}
\Phi(x) = (4N\nu-1)\ln x+(4N\mu-1)\ln(1-x) ~.
\label{Eq:2.1-10}
\end{equation}
To maintain a bi-stable system, we set $4N\nu,4N\mu<1$. There is a unique valley state (saddle point) in $(0,~1)$, here we denote as $x=a$:
\begin{equation}
a=(1-4N\nu)/(2-4N\mu-4N\nu)~,
\label{Eq:3-0}
\end{equation}
satisfying $\Phi'(a)=0$ and $\Phi''(a)>0$. The movements of a population, if visualized on the potential landscape, can be classified into two fundamentally different types: uphill and downhill processes \citep{Zhou2011}. The uphill evolution, e.g. from the saddle $x=a$ to the attractive state $x=0$, is mainly driven by the directed (adaptive) forces. By referring to the Langevin equation that describes the same evolutionary process with Eq.~(\ref{Eq:2.1-1}), we obtain the uphill rate by averaging the effects of noise over its probability distribution:
\begin{equation}
\dot{x} = f(x)=-\mu x+\nu (1-x)-(1-2x)/4N~.
\label{Eq:2.2-0}
\end{equation}
$\dot{x}$ denotes the frequency change rate; $f(x)$ is related to Eq.~(\ref{Eq:2.1-1}) by using a different stochastic integral from those of Ito and Stratonovich \citep{Ao2007}.

It is easy to verify that $\Phi$ is non-decreasing along the noise-free evolutionary trajectory of a population: $\dot{\Phi} = \Phi'(x)\cdot \dot{x}=f^2(x)/D(x)\geq 0$. For linear $f(x)$, we can always take the approximation form $f\sim -|f|x$ (here by replacing $x$ with $a-x$). The solution of Eq.~(\ref{Eq:2.2-0}) takes the approximate form $x(t)= x(0)\cdot\exp(-|f|t)\doteq x(0)\cdot\exp(-t/T_1)$, where $x(0)$ gives the initial state of the population, and $T_1$ is usually called the relaxation time. Under $4N\nu,4N\mu\ll 1$,
\begin{equation}
T_1 \sim |f|^{-1} = 2N \cdot \mathcal{O}(1) ~,
\label{Eq:2.2-1}
\end{equation}
defines the characteristic time to the local equilibria, here our first timescale.

The downhill dynamics is often considered of stochastic essence, characterized by the escape time $\tau$. Kramers' classical formula estimates
\begin{equation}
\tau \sim T_1 \exp{(\Delta \Phi)}~.
\label{Eq:2.2-4}
\end{equation}
Here $\Delta \Phi$ is the potential barrier height. However, in the present case, $\Phi(0)=\infty$, which leads to $\Delta \Phi = \Phi(0)-\Phi(a)=\infty$ and thus $\tau=\infty$. It may not be a good estimation for the escape time. Biologically, this infinity would imply gene fixation under a considerable range of forward mutation (e.g. $0<1-4N\nu \ll 1 $). Mathematically, $\tau/T_1$ would change discontinuously with $4N\nu$ (from $+\infty$ to 1) as $4N\nu\rightarrow 1$. In the rest of the work we will try to obtain better analytical approximations for $\tau$, by first referring to the mean first passage time.

\section{Downhill dynamics in infinite potential}

\subsection{Mean first passage time (MFPT)}

Without loss of generosity, we study the stochastic jump out of the attractive basin $(0,a)$. We study a population's first passage (transition) event through the valley point $x=a$ to some state $x_1> a$, starting from $x_0\approx 0$ in $(0,a)$. The mean first passage time (MFPT) $T(x_0\rightarrow x_1)$ satisfies \citep{Gardiner1985}:
\begin{equation}
T(x_0\rightarrow x_1) =\int_{x_0}^{x_1} \frac{1}{\epsilon D(y)} \exp \big[-\Phi(y)\big]dy\int_{0} ^ y \exp\big[\Phi(z)\big]dz ~.
\label{Eq:3.1-3}
\end{equation}
Here $\Phi$ is just our potential landscape in Eq.~(\ref{Eq:2.1-4}). The interested interval is set as $[0,x_1]$, with $x=0$ the reflecting boundary and $x=x_1$ the absorbing boundary. Previous approximation methods that generate the Arrehnius factor is mainly established on the following two assumptions \citep{Kramers1940, Gardiner1985}: (1) Assume a ``sharp" valley around $x=a$ on the landscape; (2) Assume Gaussian-like probability distribution around $x=0$. However, these two assumptions fail in the present model, as the landscapes typically have ``fat" valleys and sharp (even divergent) peaks under strong genetic drift (Figure 1). We seek other ways to analytically approximate the result in the present model, by first specifying Eq.~(\ref{Eq:3.1-3}) under mutation and drift:
\begin{equation}
T(x_0\rightarrow x_1) = 4N\int_{x_0}^{x_1} y^{-4N\nu}(1-y)^{-4N\mu}dy \int_0^y z^{4N\nu-1}(1-z)^{4N\mu-1}dz~. \label{Eq:3.1-4}
\end{equation}
We note that as $4N\nu,4N\mu \rightarrow 0$ the main contribution of the above integral comes from the inner integral, the incomplete Beta function $B(y;4N\nu,4N\mu)$. Under the same limit, it is numerically shown to be approximated by $1/4N\nu$. Thus the whole integral approximates a scale of $1/\nu$. More formally, we expand the incomplete Beta function in Eq.~(\ref{Eq:3.1-4}) under $0<1-x_1<1-y<1-z<1$:
\begin{equation}
B(y;4N\nu,4N\mu) = \frac{y^{4N\nu}}{4N\nu} + \sum^{\infty}_{n=1}\prod^n_{k=1}\bigg( \frac{k-4N\mu}{k} \bigg)\frac{y^{n+4N\nu}}{n+4N\nu} ~. \label{Eq:3.1-5}
\end{equation}
The convergence is obvious given $0<y<x_1<1$. Substitute $B(y;4N\nu,4N\mu)$ and expand $(1-y)^{-4N\mu}$ in the outer integral of Eq.~(\ref{Eq:3.1-4}), we obtain
\begin{align}
&T(x_0\rightarrow x_1) = \dfrac{x_1-x_0}{\nu} + \dfrac{1}{\nu}\bigg[2N\mu(x_1^{2}-x_0^{2}) + 4N\mu\sum^{\infty}_{n=2}\prod^n_{k=2}\bigg( \frac{k-1+4N\mu}{k} \bigg) \dfrac{x_1^{n+1}-x_0^{n+1}}{n+1}\bigg] ~ + \notag \\
&~~~~~~~~ 4N\bigg[\dfrac{1-4N\mu}{1+4N\nu}\dfrac{x_1^{2}-x_0^{2}}{2} + (1-4N\mu)\sum^{\infty}_{n=2}\prod^n_{k=2}\bigg( \frac{k-4N\mu}{k} \bigg)\dfrac{x_1^{n+1}-x_0^{n+1}}{(n+1)(n+4N\nu)} \bigg] ~.
\label{Eq:3.1-6}
\end{align}
The convergence of this expansion is obvious under $\nu>0,~\mu<1/4N$. For

(1) $\nu\rightarrow 0$: The expansion of Eq.~(\ref{Eq:3.1-5}) becomes invalid. The leading term of the expansion changes from $y^4N\nu/4N\nu$ to $\ln y$, which becomes sensitive to $x_0$ near 0 then. To ensure the convergence of $T(x_0\rightarrow x_1)$ as $x_0\rightarrow 0$, we need $\nu\neq 0$; this is the condition for the escape problem (from $x=0$) to be well-defined. On the other hand, we always have $T(0\rightarrow x_1)\rightarrow \infty$ as $\nu\rightarrow 0$.

(2) $\mu\rightarrow 1/4N$: The expansion of $(1-y)^{-4N\mu}$ would not converge for $x_1\rightarrow 1$, as the resulted series would then becomes a divergent harmonic series. This is also illustrated by the vanishing bi-stability of the system (Figure 1, yellow). To ensure the convergence of Eq.~(\ref{Eq:3.1-6}) as $x_1\rightarrow 1$, we need $\mu<1/4N$.

\subsection{Escape time}

\cite{Kramers1940} first calculated the stationary flux rate of probability as the escape rate from the attractive basin. Its equivalence to the MFPT was discussed by \cite{Hanggi1990}, and should be compensated by a factor of 2 if we choose the saddle $x_1=a$ as a perfect absorbing boundary (sink) rather than a smooth distribution of sinks in $(a,1)$. Under $4N\nu,4N\mu\ll 1$, we have by Eq.~(\ref{Eq:3-0}) that $a=1/2$, and the escape time $\tau_0$ reads then (take $x_0=0$):
\begin{align}
\tau_0&= 2\times T(0\rightarrow 1/2) \notag \\
&= \dfrac{1}{\nu} + \dfrac{1}{\nu}\bigg[N\mu + 4N\mu\sum^{\infty}_{n=2}\prod^n_{k=2}\bigg( \frac{k-1+4N\mu}{k} \bigg) \dfrac{2^{-n}}{(n+1)}\bigg] ~ + \notag \\
&\qquad 4N\bigg[\dfrac{1-4N\mu}{4(1+4N\nu)} + (1-4N\mu)\sum^{\infty}_{n=2}\prod^n_{k=2}\bigg( \frac{k-4N\mu}{k} \bigg)\dfrac{2^{-n}}{(n+1)(n+4N\nu)} \bigg] ~.
\label{Eq:3.2-1a}
\end{align}
Even though the validity of MFPT in Eq.~(\ref{Eq:3.1-6}) does not rely on the existence of a potential landscape, its equivalence to the escape time requires this concept: $(0,a)$ should be a domain of attraction. And under $4N\nu,4N\mu\ll 1$, the escape time is approximately independent of the initial state $x_0$ (source) in Eq.~(\ref{Eq:3.2-1a}):
\begin{equation}
\tau_0 \approx \nu^{-1}(1+1.23N\mu)~, \label{Eq:3.3-7a}
\end{equation}
The coefficient 1.23 is an approximation of the series in Eq.~(\ref{Eq:3.2-1a}). $\tau_0$ is much bigger than the relaxation time $2N$ in Eq.~(\ref{Eq:2.2-1}). This shows the separation of the two timescales and completes our inquiry for the escape time from $(0,a)$.

Another way to look at the MFPT in Eq.~(\ref{Eq:3.1-6}) is to set $x_1=1$ and obtain the substitution time of $A_1$ alleles. It differs from the escape time above by taking into account the dynamical details in the other attractive basin $(a,1)$:
\begin{align}
T(0\rightarrow 1)&= \dfrac{1}{\nu} + \dfrac{1}{\nu}\bigg[2N\mu + 4N\mu\sum^{\infty}_{n=2}\prod^n_{k=2}\bigg( \frac{k-1+4N\mu}{k} \bigg) \dfrac{1}{n+1}\bigg] ~ + \notag \\
&\qquad 4N\bigg[\dfrac{1-4N\mu}{2(1+4N\nu)} + (1-4N\mu)\sum^{\infty}_{n=2}\prod^n_{k=2}\bigg( \frac{k-4N\mu}{k} \bigg)\dfrac{1}{(n+1)(n+4N\nu)} \bigg] ~.
\label{Eq:3.2-1b}
\end{align}
The necessary condition for its convergence ($\nu> 0,~\mu<1/4N$) has been discussed in Section 3.1. In Appendix we show that the condition is also sufficient. In general we see that $T(0\rightarrow 1)>\tau_0$, as the backward mutation would become much stronger for $x\in(a,1)$. Under the limit $4N\nu,~4N\mu\ll 1$, the two equations arrive at the same result $T(0\rightarrow 1)\approx \tau_0\approx 1/\nu$. Numerical comparison of Eq.~(\ref{Eq:3.1-4}) (take $x_0=0,~x_1=1$), Eq.~(\ref{Eq:3.3-7a}) and results from the discrete model \citep{Blythe2007} is given in Figure 2(a).

Escape time $\tau_1$ from the attractive basin $(a,1)$ can be similarly derived like above. Transitions in the two directions will eventually balance each other, and the global equilibrium distribution will be established in the second timescale (inverse of the leading flux rate toward equilibrium \citep{Hanggi1990}):
\begin{equation}
T_2\sim \Big(\dfrac{1}{\tau_0}+\dfrac{1}{\tau_1}\Big)^{-1} = \dfrac{\tau_0\tau_1}{\tau_0+\tau_1} \approx \dfrac{1}{\mu+\nu} ~. \notag
\end{equation}
Simulation of the dynamical behaviors of the discrete Wright-Fisher model in the two timescales is shown in Figure 3.

\subsection{Effect of weak selection}
The general equation for the escape time when there is mutation, drift and selection is obtained by substituting Eq.(\ref{Eq:2.1-9}) into Eq.(\ref{Eq:3.1-3})
\begin{equation}
T(x_0\rightarrow x_1) = 4N\int_{x_0}^{x_1} (1-y)^{-4N\mu}y^{-4N\nu} \big[\overline{\omega}(y)\big]^{-2N} dy \int_0^y (1-z)^{4N\mu-1}z^{4N\nu-1} \big[\overline{\omega}(z)\big]^{2N} dz~.
\label{Eq:3.3-1}
\end{equation}
If we can expand $[\overline{\omega}(y)]^{-2N}$ and $[\overline{\omega}(z)]^{2N}$ near 0, an analytical approximation for $T(0\rightarrow 1)$ is obtained by combining the results with Eq.~(\ref{Eq:3.2-1b}). For example, if we take $s$ as the selective advantage of $A_1$ over $A_2$ ($s\ll 1$), such that the average rate of change in $x$ per generation by selection is \citep{Kimura1964} $M_s=sx(1-x)$, the average fitness is given by $\overline{w}= 1+2sx~$. To maintain a bi-stable system, we set $1/4N>\mu,\nu$. To take the expansion we further assume $4Ns<1$. Substitute above settings into Eq.~(\ref{Eq:3.3-1}) and obtain:
\begin{align}
T(0\rightarrow 1) \approx (1+2N\mu-2Ns)/\nu ~, \label{Eq:3.3-3}
\end{align}
the substitution time of $A_1$ alleles. From this result, the selective advantage $s$ decreases the substitution time approximately on a linear scale if $4Ns<1$, consistent to the rate of substitution calculated under the same settings without backward mutations ($\mu=0,s\ll 1, 4Ns<1$) \citep{Gillespie1998}:
\begin{equation}
k = \dfrac{1-e^{-2s}}{1-e^{-4Ns}} \times 2N\nu \approx \dfrac{\nu}{1-2Ns}~, \notag
\end{equation}
just the inverse of Eq.(\ref{Eq:3.3-3}) if we take $\mu=0$. Another example of selection is described by \cite{Barton1987}, taking the form $M_s(x)=-sx(1-x)(1-2x)$ and $\mu=\nu$, where $s$ is the fitness deficit of the heterozygote relative to the homozygote. We have $\overline{w}= 1-2sx+2sx^2$. The potential landscape is
\begin{equation}
\Phi(x) = (4N\mu-1)\ln x(1-x)-4Nsx+4Nsx^2 ~,
\label{Eq:3.3-6}
\end{equation}
plotted in Figure 1 (cyan). The peak-to-peak transition rate (which \cite{Barton1987} failed to approximate) is then obtained as
\begin{align}
\tau_0^{-1} = 1/(2\times T(0\rightarrow a)) \approx \mu/(1+1.23N\mu+0.67Ns)~. \label{Eq:3.3-7b}
\end{align}
Here $a=1/2$. We give numerical comparisons among Eq.~(\ref{Eq:3.3-1}) (specified by Eq.~(\ref{Eq:3.3-6}) and take the inverse), Eq.~(\ref{Eq:3.3-7b}) and discrete results in Figure 2(b).

\section{Discussion}
\subsection{Comparisons with previous work}
From Eq.~(\ref{Eq:3.3-7a}), the transition time is approximately independent of the population size $N$. It reminds us of Kimura's famous rate formula for the neutral evolution \citep{Kimura1962}, or the rate of neutral substitution: $2N\nu \times 1/2N = \nu$, just the inverse of Eq.~(\ref{Eq:3.3-7a}) if we take $\mu=0$. The coincidence happens under the limit $4N\nu\ll 1$; for comparable $\nu$ and $1/2N$, however, the population size $N$ will have significant effects on the transition rate, and this simple estimation will fail. Our Eq.~(\ref{Eq:3.2-1a}) instead gives a more general rate formula for the neutral evolution. It allows the existence of other types of biological factors, e.g. the two-way mutations, which may make the fixation probability of a new mutant (and thus the rate of substitution) incalculable. Our results in Section 3.3 may also help test the neutrality of specific biological systems.

Our results show that Kramers' classical escape formula can be extended to the non-Gaussian distribution cases. Under $4N\nu,~4N\mu\ll 1$, the result does not show exponential dependency on the valley depth (or barrier height), but rather is controlled by the sharpness of the potential peak (see the sensitivities of Eq.~(\ref{Eq:2.1-10}) and Eq.~(\ref{Eq:3.3-7a}) with respect to $\nu$). On the other hand, under $4N\nu,~4N\mu\ll 1$, we have $T_1\sim 2N$ and $T_2\sim \nu^{-1}$; there is still the separation of different timescales, which naturally emerges from our expansion Eq.~(\ref{Eq:3.2-1a}).

\citet{Barton1987} use the eigenvalue method to study the second example in our Section 3.3. Their method failed to approximate the transition rate under very weak selection ($s<4\mu$), however, as the approach requires two peaks on their ``deterministic equilibrium". This requirement is not satisfied in many systems that show long-term bi-stability, e.g. the mutation-drift and the weak selection cases discussed in Section 3. Also, part of their solution cannot be analytically expressed except for some limiting cases. In comparison, our Eq.~(\ref{Eq:3.1-6}) directly expands the non-Gaussian (divergent) equilibrium distribution, and our results can be used for the whole parameter subspace which maintains system's long-term bi-stability in the present model.

\subsection{More on our potential landscape}
As shown in Section 2.1, $\Phi$ relates to $\rho(x,~t=\infty)$ through the Boltzmann-Gibbs distribution if $Z<\infty$. If otherwise $Z=\infty$, the definition in Eq.~(\ref{Eq:2.1-4}) is still valid and changes continuously with the parameters of the system. An example is to take $\nu=\mu=0$ in Eq.~(\ref{Eq:2.1-10}), meanwhile the stationary distribution becomes a combination of the Dirac delta functions $\rho(x,~\infty)=C\delta(x) + (1-C)\delta(1-x)~.$
$C$ is a constant depending on the initial system state \citep{Mckane2007}. We plot $\Phi$ and $\rho(x,~\infty)$ in Figure 1 (red).

The potential landscape Eq.~(\ref{Eq:2.1-4}) can be compared to the classical fitness landscape, which presents only the effects of selection. Other biological factors may generate various evolutionary mechanisms on the fitness landscape without a unified description, along with other controversies \citep{Kaplan2008}. Also, by only taking the measure of fitness, there may be inconsistencies between the dynamics and biology. One example is the term ``neutral evolution" commonly used in the absence of selection, where different allele-frequency states of a population are not necessarily equally favored by evolution (except the special case $4N\nu=4N\mu=1$), shown in Figure 1. The present potential landscape may serve as a substitute for \cite{Wright1932}'s original landscape that visualizes and quantifies the evolutionary process in a globally coherent way.

An extension to the fitness landscape is the so-called ``deterministic equilibrium" \citep{Barton1987}, which integrates all other factors except random drift. It fails to capture the bi-stability of the system when the stochastic effect has a non-trivial contribution to the evolution direction. The associated approaches also fail for such cases (see Sections 4.1 and 3.3). Another extension is the free fitness function used by \citet{Barton2009}, in consideration of the analogy with thermodynamics. But there are cases (e.g. weak mutation) where their maximum entropy approximation fails, and their method ``assumes normality". Our framework does not have certain constraints, and the validity of our landscape construction and the associated approaches is tested in the whole relevant parameter space. It has already been applied in the study of Muller's ratchet \citep{Jiao2012}, a special case where no backward mutations exist.

\subsection{Normalization constants and fixation}

By taking $\nu=0$ in Eq.~(\ref{Eq:3.1-6}), we have $\tau_0=+\infty$. No escape is expected to happen once a population ``trapped" into the neighborhood of $x=0$. In Eq.~(\ref{Eq:3.1-3}), the impossibility comes essentially from the infinity of $B(y;4N\nu,4N\mu)$ in Eq.~(\ref{Eq:3.1-5}). More formally, if we define a partial normalization constant for each attractive basin (taking the mutation-drift case as an example) as
\begin{align}
Z_1 = \int^a_0 x^{4N\nu-1}(1-x)^{4N\mu-1} dx~, ~ Z_2 = \int^1_a x^{4N\nu-1}(1-x)^{4N\mu-1} dx~, \label{Eq:4.3-2}
\end{align}
then the mathematical condition for the biological fixation at $x=0$ (or $x=1$) should be $Z_1=\infty$ (or $Z_2=\infty$), not $\Phi(0)=+\infty$ (or $\Phi(1)=+\infty$). We call it ``complete fixation" if populations starting from any initial state will finally be fixed at a monomorphic gene state $x=0$ or $x=1$, here determined by $Z_1=\infty,~ Z_2<\infty$ or $Z_1<\infty,~ Z_2=\infty$. If $Z_1=Z_2=\infty$, the fixation will happen at either $x=0$ or $x=1$ on probability. Another observation from the results is the emerging of absorbing boundaries at the fixation state; the boundary conditions ``artifically" set by \citet{Mckane2007} are more naturally and generally derived then. Our last comment is that unnormalizable distributions in the diffusion model do not generate real problems, but instead provide important dynamical and equilibrium information for the understanding of the system. We summarize above conclusions in Table 1.

\begin{table}[h]
\centering
\caption{Summary of the observations in Section 4.3. $Z_1$ and $Z_2$ are the partial normalization constants defined in Eq.~(\ref{Eq:4.3-2}). ``Complete" fixation and ``Fix on prob." are fixation types defined in Section 4.3. $\tau_0$ and $\tau_1$ are the respective escape times. The ``Absorb-bound." column gives where the absorbing boundary emerges.}
\begin{tabular}{c c c c c c c}
\hline
$Z_1$ & $Z_2$ & $\tau_0$ & $\tau_1$ & Fixation type & Absorb-bound.
\\ \hline
$<\infty$ & $<\infty$ & $<\infty$ & $<\infty$ & N/A & Neither
\\ \hline
$=\infty$ & $<\infty$ & $=\infty$ & $<\infty$ & Complete~($x=0$) & $x=0$
\\ \hline
$<\infty$ & $=\infty$ & $<\infty$ & $=\infty$ & Complete~($x=1$) & $x=1$
\\ \hline
$=\infty$ & $=\infty$ & $=\infty$ & $=\infty$& Fix on prob.  & $x=0,1$
\\ \hline
\end{tabular}
\end{table}

\subsection{Comments on the ``stochastic tunneling"}

\citet{Iwasa2004} studied a three-phase transition problem, in which they termed a ``stochastic tunneling" phenomenon that allows transition from one state to another, without passing through the middle state. Here our first comment is that their use of the term implies the existence of a potential barrier (or adaptive valley), and thus a landscape. Second, the term ``tunneling" is misleading as it refers to some quantum dynamics which is classically impossible. The actually process is driven by noise and should be properly described as climbing over a saddle point on a 2-d landscape surface.

\section*{Acknowledgements}
The critical comments of D. Waxman and T. Kr\"{u}ger on this work are appreciated. We also thank R. S. Yuan, J. H. Shi, Y. B. Wang, and other members in the lab for their constructive comments. We thank X. A. Wang for the technical support. This work was supported in part by the National 973 Project No.~2010CB529200; and by the Natural Science Foundation of China No.~NFSC61073087 and No.~NFSC91029738.

\bibliographystyle{plain}
\bibliography{JMB}

\section*{Appendix}
\setcounter{equation}{0}
\renewcommand{\theequation}{A.\arabic{equation}}
Under $\nu>0$, the convergence of Eq.~(\ref{Eq:3.2-1b}) relies on the convergence of the sum
\begin{equation}
S = \sum^{\infty}_{n=2}\prod^n_{k=2}\bigg( \frac{k-1+4N\mu}{k} \bigg) \frac{1}{n+1} ~. \notag
\end{equation}
We use Raabe's test for series convergence from standard textbooks of real analysis. For $0\leq 4N\mu<1$, we denote
\begin{equation}
c_n = \prod^n_{k=2}\bigg( \frac{k-1+4N\mu}{k} \bigg) \frac{1}{n+1}~. \notag
\end{equation}
Obviously $c_n$ is positive for all $n>0$. First, we have
\begin{equation}
\lim_{n\rightarrow\infty}\frac{c_{n+1}}{c_n}=1~.
\label{Eq:A2-1}
\end{equation}
We then calculate the Raabe terms
\begin{equation}
R_n = n\bigg( \frac{c_{n+1}}{c_n}-1 \bigg) = \big(4N\mu -2\big)\frac{n}{n+2} ~. \notag
\end{equation}
Here $4N\mu-2$ is a constant less than $-1$. By taking the limit $n\rightarrow \infty$,
\begin{equation}
\lim_{n\rightarrow \infty}R_n = 4N\mu -2 < -1
\label{Eq:A2-2}
\end{equation}
The two conclusions in Eqs.(\ref{Eq:A2-1}, \ref{Eq:A2-2}) verify the convergence of the partial sum $S_n$ under $0\leq 4N\mu<1$.

\begin{figure}[h]
\centering
\subfigure[Potential landscapes]{
\includegraphics[width=112mm, height=84mm]{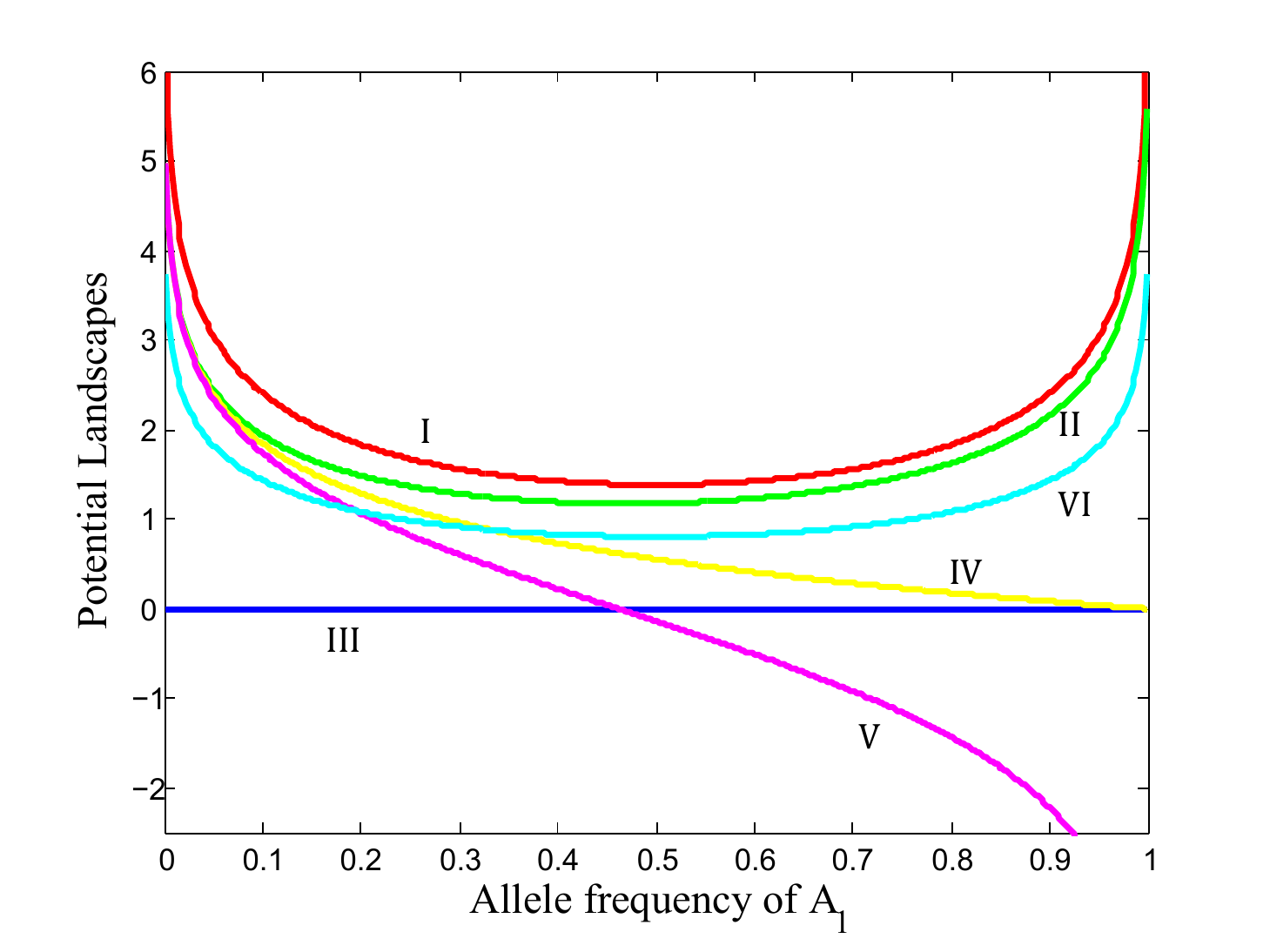}
}
\subfigure[Equilibrium distributions]{
\includegraphics[width=112mm, height=84mm]{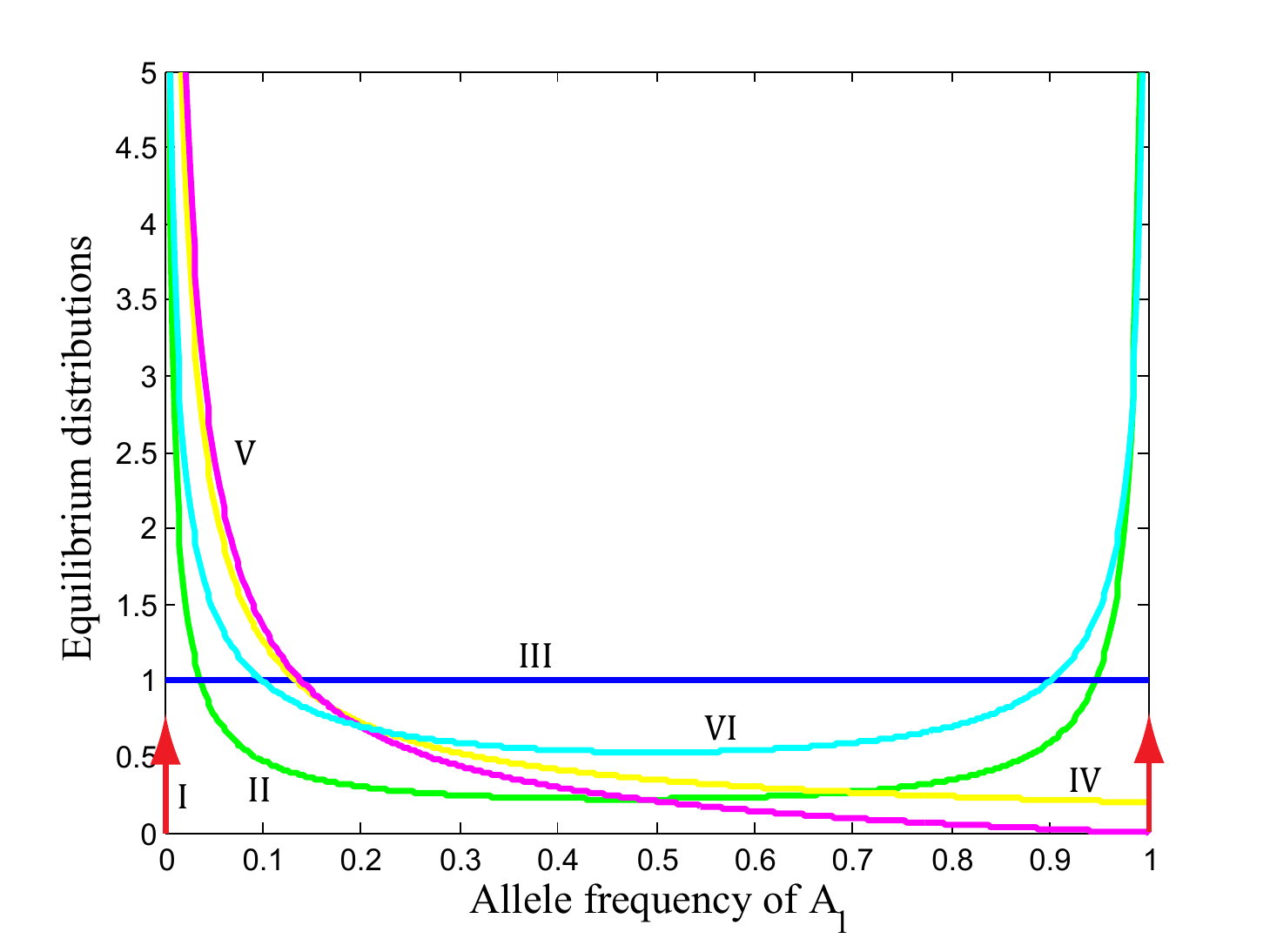}
}
\caption{Potential landscapes and corresponding equilibrium distributions under different parameter settings in the Wright-Fisher model, differentiated by both the colors and Roman indexes. In all cases there is $N=50$. The following five colored landscape contours are generated from Eq.~(\ref{Eq:2.1-10}) under mutation and genetic drift: Red (I): $\mu=\nu=0$. Green (II): $\mu=0.0005,~\nu=0.001$. Blue (III): $\mu=\nu=0.005$. Yellow (IV): $\mu=0.005,~\nu=0.001$. Magenta (V): $\mu=0.01,~\nu=0.001$. The last one is generated from Eq.~(\ref{Eq:3.3-6}), considering mutation, drift, and selection: Cyan (VI): $\mu=\nu=0.002,~s=0.1$. The two red arrows in (b) denote the Dirac delta functions.}
\end{figure}

\begin{figure}[h]
\centering
\subfigure[Mutation and genetic drift]{
\includegraphics[width=112mm, height=84mm]{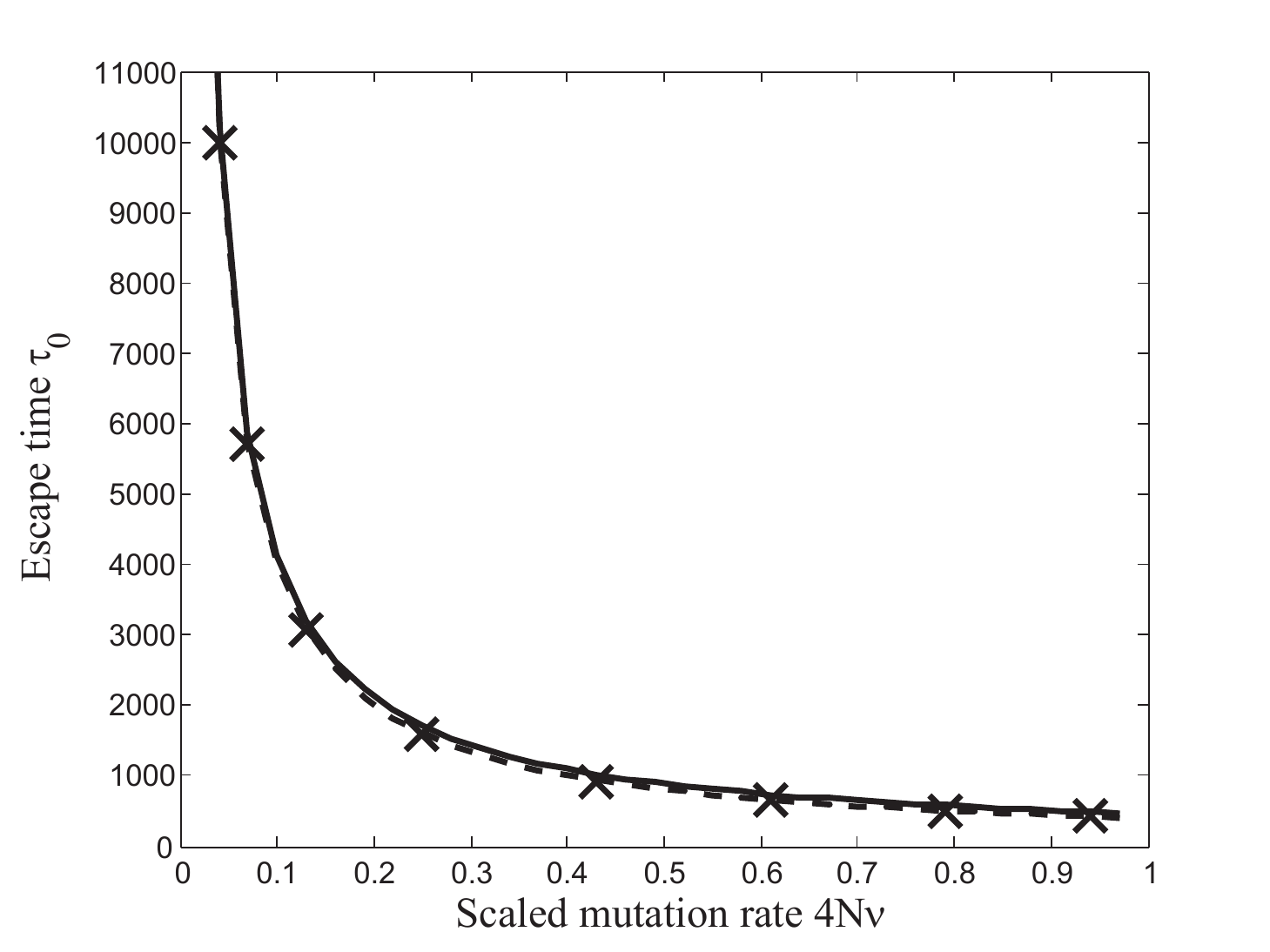}
}
\subfigure[Selection, mutation and genetic drift]{
\includegraphics[width=112mm, height=84mm]{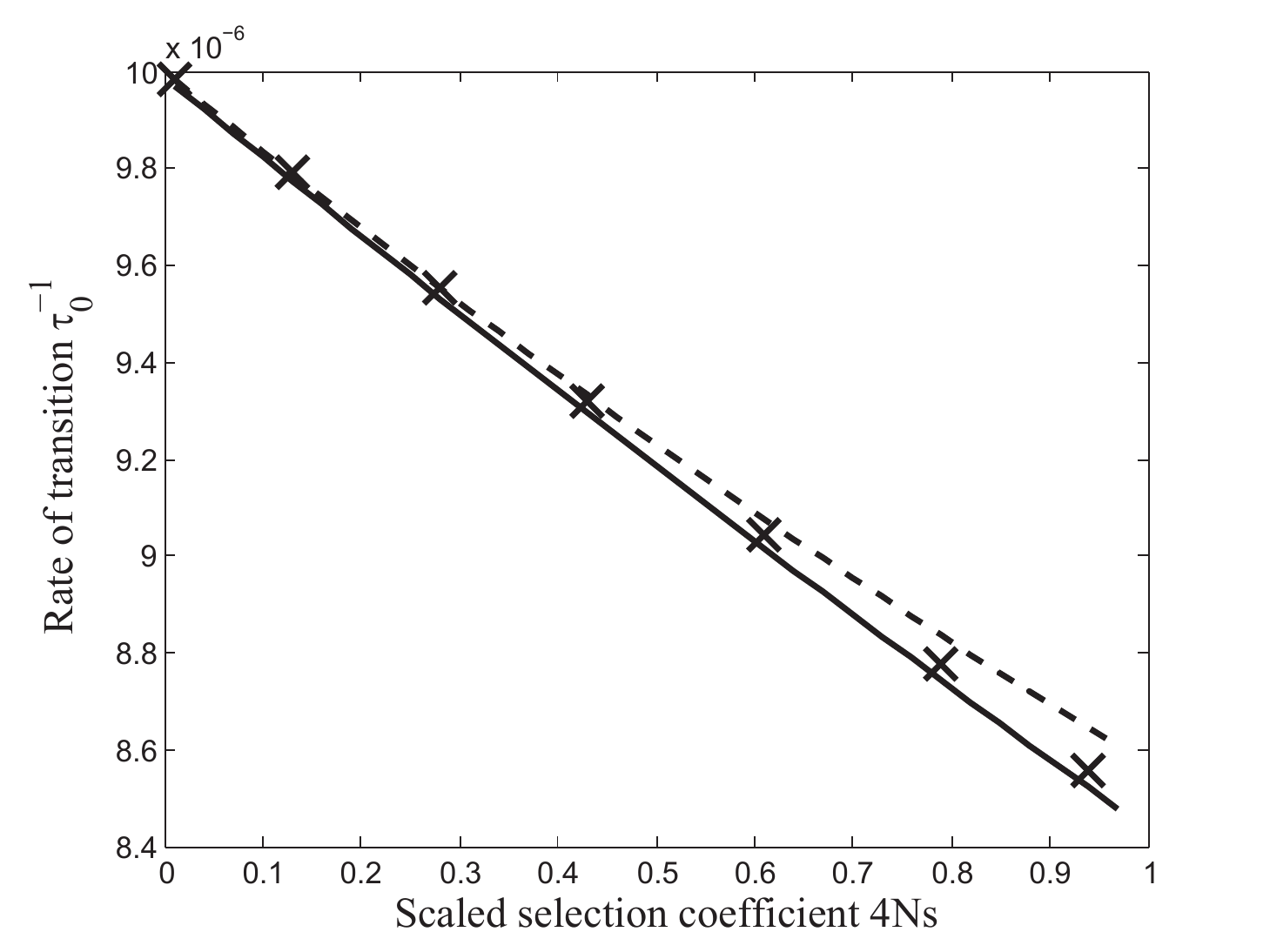}
}
\caption{Analytical approximations (dashed) of the escape time/rate compared with the numerical solutions (solid) and results in the discrete Wright-Fisher model (numerical calculations of the first non-vanishing eigenvalues of the transition probability matrix, denoted by crosses). (a) Mutation and random drift ($N=100$). The escape times are compared. The range of the x-axis is chosen in a way that system's bi-stability is maintained. (b) Selection, mutation and random drift ($N=50$). The rates of escape are compared. The range of x-axis is chosen so that weak selections are considered.}
\end{figure}

\begin{figure}[h]
\centering
\subfigure[t=0]{
\includegraphics[width=56mm, height=50mm]{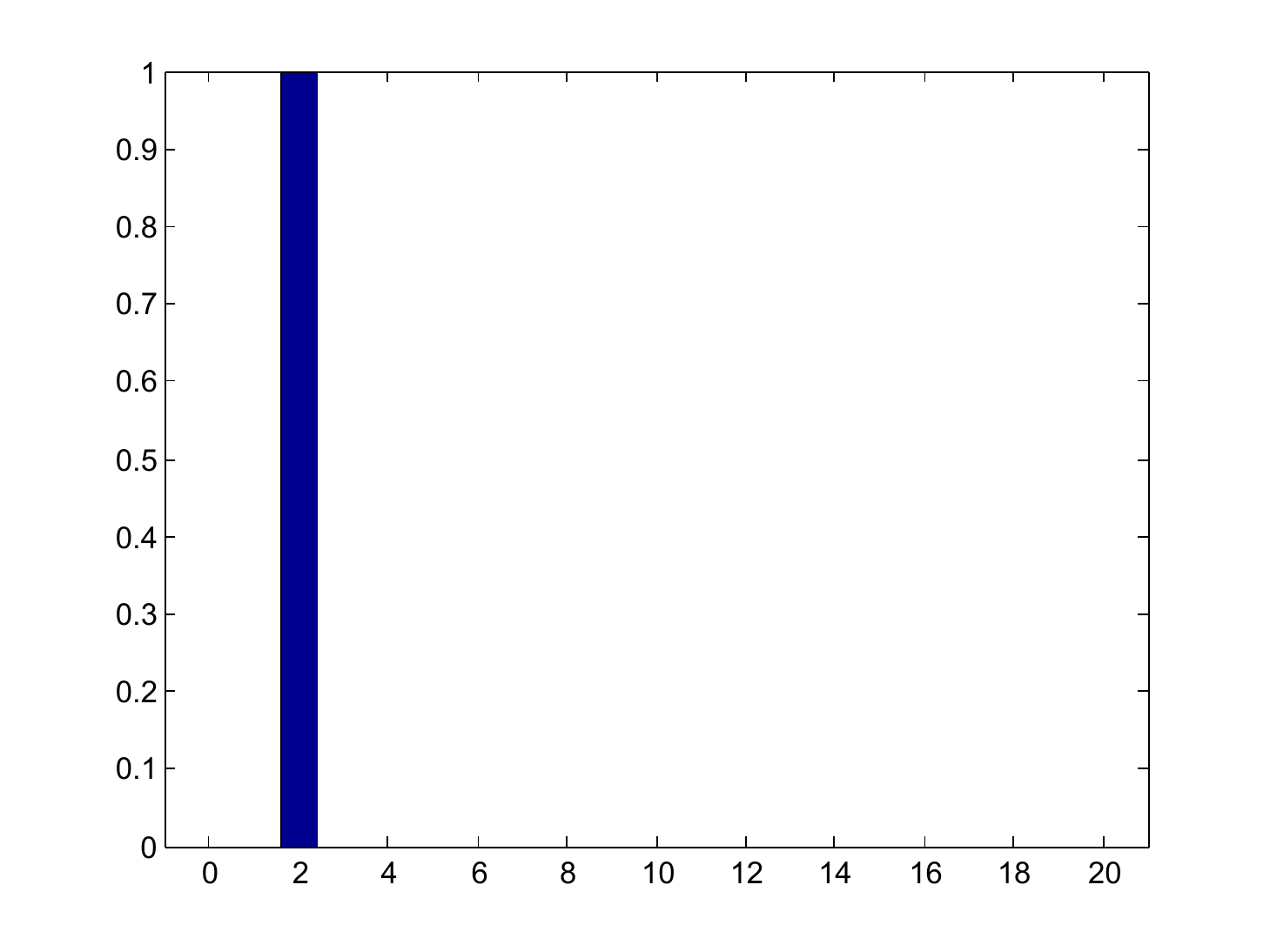}
}
\subfigure[t=20]{
\includegraphics[width=56mm, height=50mm]{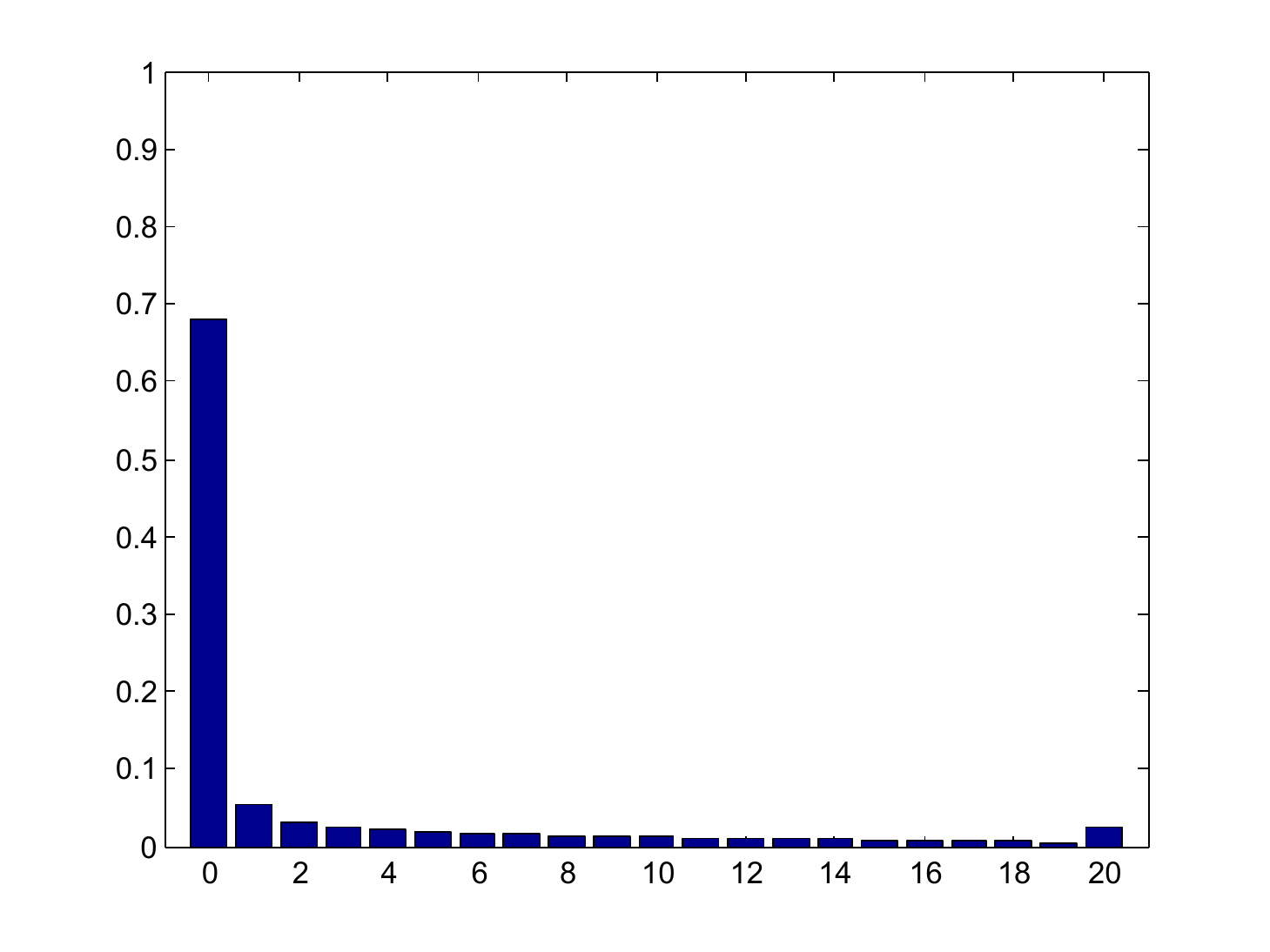}
}
\subfigure[t=666]{
\includegraphics[width=56mm, height=50mm]{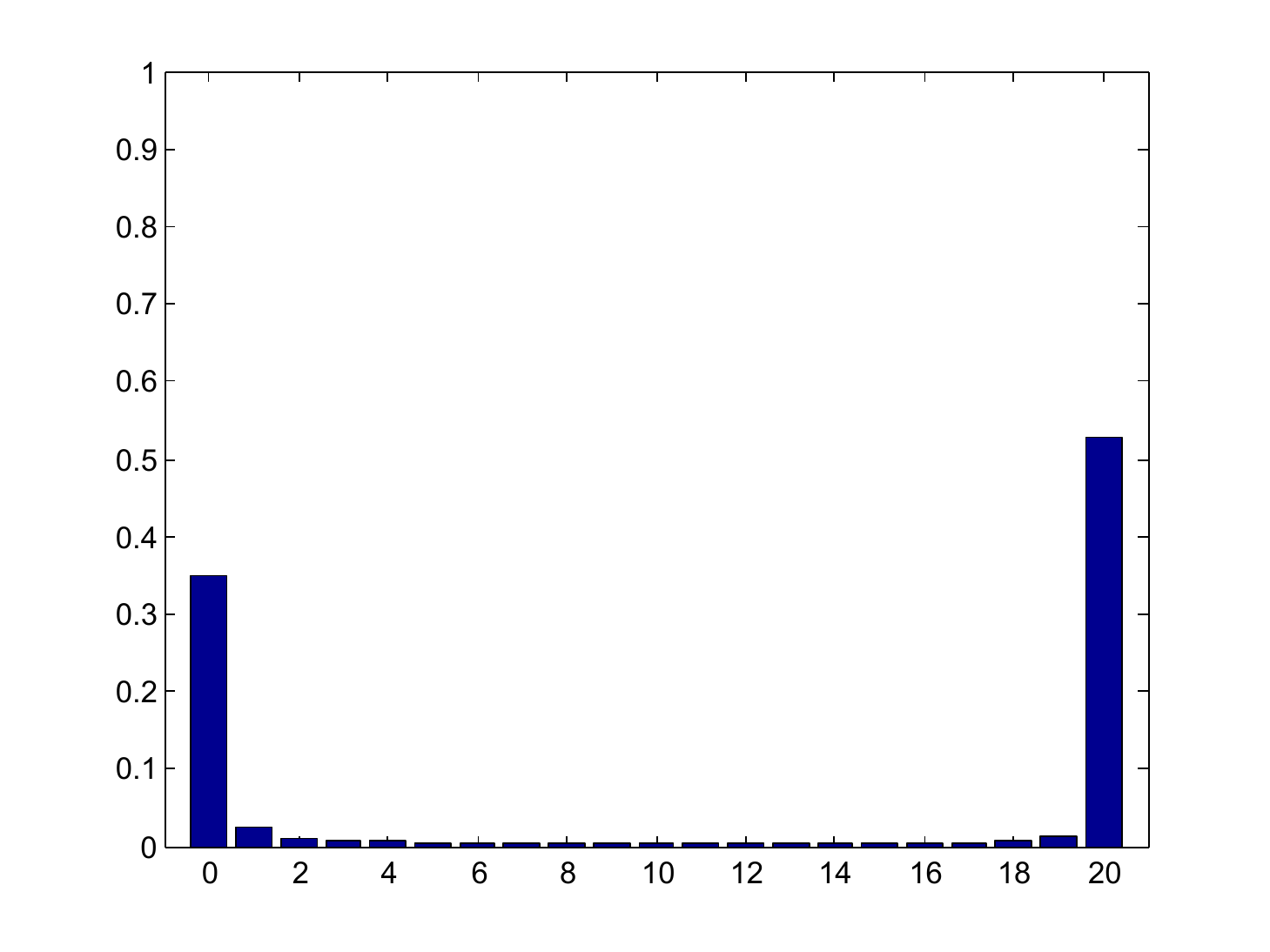}
}
\subfigure[t=1500]{
\includegraphics[width=56mm, height=50mm]{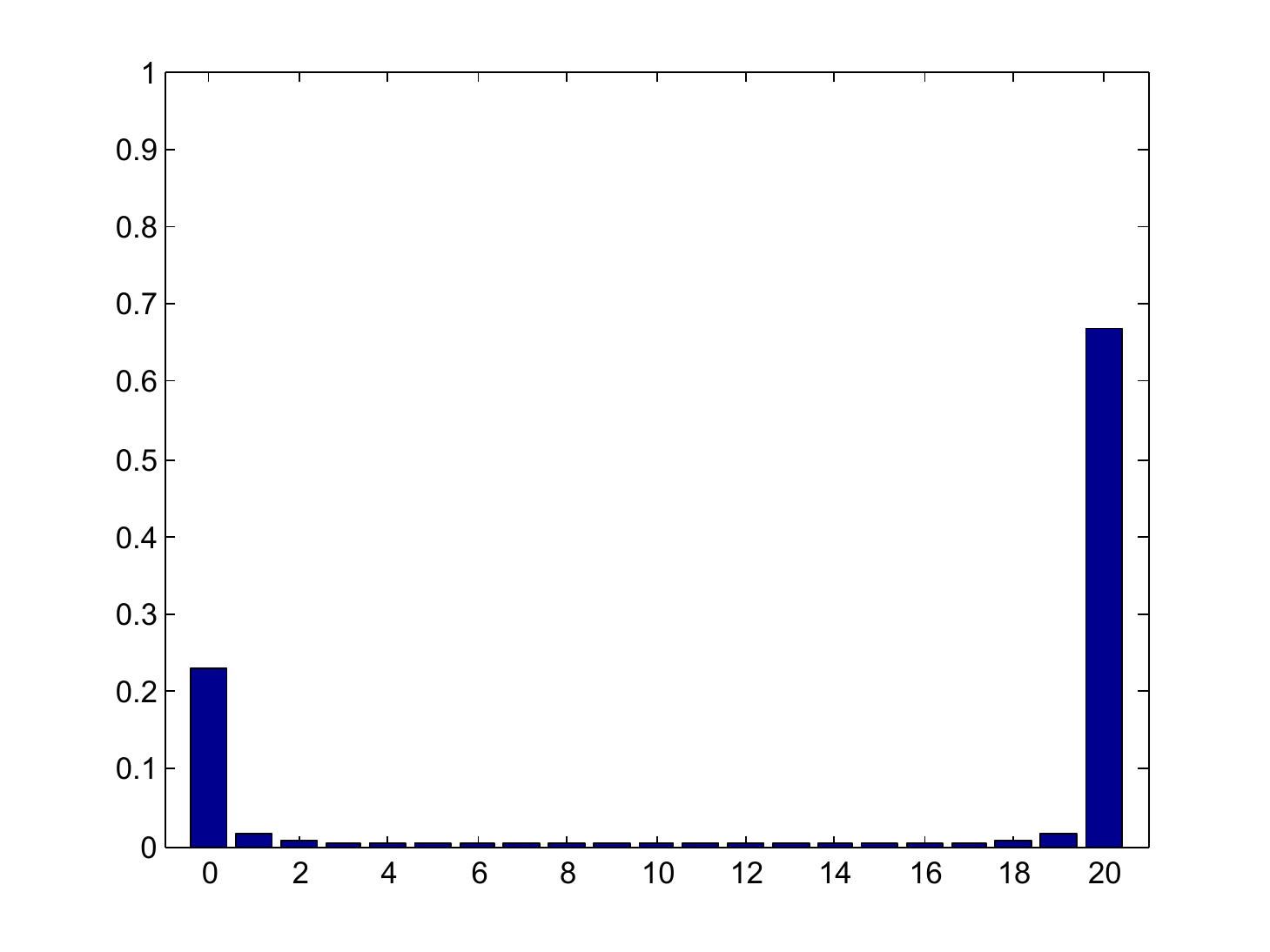}
}
\subfigure[Landscape visualization]{
\includegraphics[width=112mm, height=84mm]{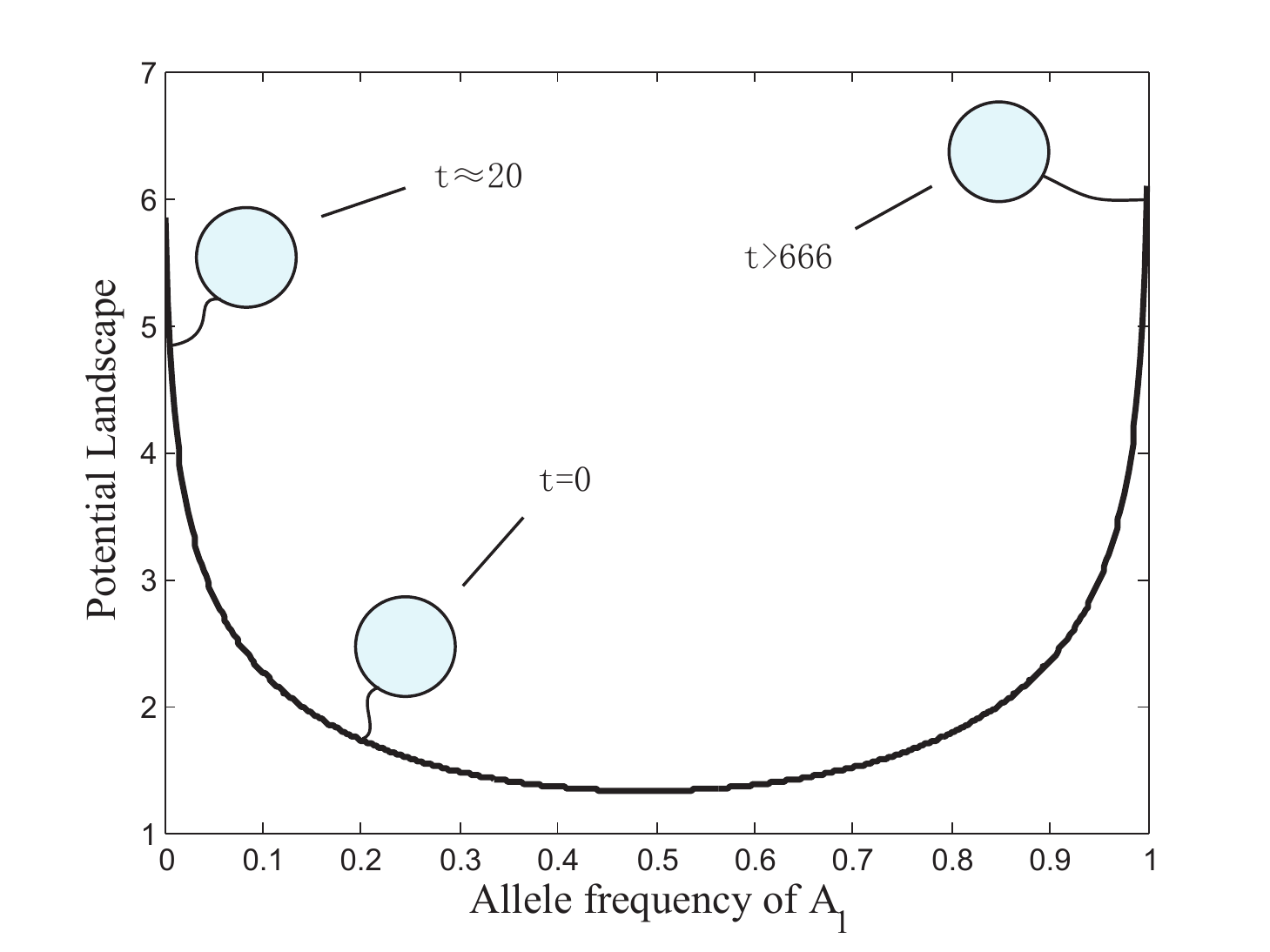}
}
\caption{Simulations realized from the discrete Wright-Fisher model under mutation and random drift. In (a)-(d), x-axis gives the number of $A_1$ alleles and y-axis is the probability distribution. Parameter settings: $2N=20,~\mu=0.0005,~\nu=0.0015$, so that $T_1\approx20,~\tau_0\approx 666$. (a) shows that the initial state is set to $x=0.2$. (d) shows the establishment of the equilibrium distribution after long enough time. (e) gives the most probable state of a population (denoted as a balloon, which always searches for a higher ``altitude" to stay) in different timescales visualized on the potential landscape.}
\end{figure}

\end{document}